\documentclass[conference,a4paper]{IEEEtran}
\IEEEoverridecommandlockouts

\usepackage{balance}

\usepackage{amsmath}
\usepackage{amssymb}
\usepackage{graphicx}

\usepackage{mathtools}

\title{Aspects of density approximation by tensor trains\\
\thanks{This work was supported by the Czech Science Foundation, project no.~P103/22-11101S, and by the Ministry of Education, Youth and Sports of the Czech Republic under project ROBOPROX -- Robotics and Advanced Industrial Production CZ.02.01.01/00/22\_008/0004590.}
}

\author{\IEEEauthorblockN{Ji\v{r}\'{i}~Ajgl, Ond\v{r}ej~Straka}
\IEEEauthorblockA{\textit{European Centre of Excellence -- New Technologies for Information Society and}\\ \textit{Department of Cybernetics,
Faculty of Applied Sciences, University of West Bohemia in Pilsen,}\\
Pilsen, Czech Republic.\\
jiriajgl@kky.zcu.cz, straka30@kky.zcu.cz}}

\begin{document}

\maketitle

\begin{abstract}
 Point-mass filters solve Bayesian recursive relations by approximating probability density functions of a system state over grids of discrete points. The approach suffers from the curse of dimensionality. The exponential increase of the number of the grid points can be mitigated by application of low-rank approximations of multidimensional arrays. Tensor train decompositions represent individual values by the product of matrices. This paper focuses on selected issues that are substantial in state estimation. Namely, the contamination of the density approximations by negative values is discussed first. Functional decompositions of quadratic functions are compared with decompositions of discretised Gaussian densities next. In particular, the connection of correlation with tensor train ranks is explored. Last, the consequences of interpolating the density values from one grid to a new grid are analysed. 
\end{abstract}


\section{Introduction}
 Bayesian filtering \cite{Sarkka:23} calculates the posterior density of the state of a stochastic system. Since the functional problem is intractable in a general non-linear non-Gaussian case, various approximations were developed. Kalman type filters focus on moments and are frequently presented as founded on the assumption of gaussianity. The fundamental idea is to process a small number of parameters with an intuitive meaning, such as central tendency and spread of the uncertainty. On the opposite side of the full spectrum of filters, there are classical non-parametric filters. A huge number of discrete objects is used to represent a probability density function, where the individual items bear no specific interpretation. Particle filters \cite{Gustafsson:10,Godsill:19} generate random samples and provides them with weights. The collection of samples is not organised and the density is not readily available. Point mass filters \cite{Simandl:06}\nocite{Dunik:19,Matousek:23}--\cite{Matousek:25} 
design regular grids in each time instant and represent the density values directly. The standard form is a matrix or a three-dimensional array, since the exponential increase in computational requirements prevents the filters from being used for multi-dimensional problems.

 Multilinear algebra offers tools for dealing with higher-dimensional arrays, which are called tensors. Tensor decompositions \cite{Rabanser:17,Grasedyck:13} can be used to construct low-rank approximations. Various approaches exist under different names. The basic approaches generalise the singular value decomposition of matrices. The canonical polyadic decomposition considers the sum of outer products of individual vectors. The Tucker decomposition scales a core tensor by multiple matrices, whereas the canonical decomposition is obtained by restricting the core to be "diagonal". The tensor train decomposition \cite{Oseledets:11,Oseledets:TTT} is based on matrix products, where the order of the factors is an important design parameter.
 
 The low-rank approximations of tensors are valuable when full tensors have to be stored in a memory, e.g.\ when one wishes to store a trained neural network in a mobile device. However, there are other aspects. The basic algorithms like the alternating least squares or the sequential application of the singular value decomposition access all values of the arrays, which implies large memory requirements during the construction of the approximate tensors. Techniques avoiding such requirements are available. The pseudoskeleton factorisation of matrices \cite{Goreinov:97} is employed in cross interpolations of tensors \cite{Savostyanov:14,Tichavsky:24}. Cross algorithms treat full tensors as functions that can be evaluated in arbitrary indices. They use a stochastic search and iteratively expand the factors exclusively via the elements which have been evaluated so far. The approximate tensors may be processed directly in the decomposed form, although few point-wise functions of tensors can be handled analytically \cite{Litvinenko:23}.

 Instead of dealing with arrays of numerical values, decomposition of multivariate functions defined on a continuous space can be considered \cite{Bigoni:16,Gorodetsky:19}. The continuous range is standard in point mass filtering anyway. The link with the discrete arrays is that these arrays contain weights of basis functions or that the arrays contain values sampled at discrete points. The latter approach is followed in this paper. Functional tensor trains replace the product of index-dependent matrices by a product of matrix-valued functions of scalar arguments. The sizes of these matrices drive the computational requirements. Some functions can be decomposed analytically with low ranks \cite{Oseledets:13}. The functional decomposition has been applied to the transient density in \cite{Tichavsky:23}, the low-rank approximation can cope with a performance bottleneck in Bayesian filtering.

 Representation of state densities by tensors has been studied in diverse works. The canonical decomposition and the Fokker-Planck equations are discussed in \cite{Sun:15,Govaers:19}. The papers show proofs of the concept, but offer no persuasive examples for higher dimension. Various aspects of filtering and smoothing are explored in \cite{Zhao:23}. Rank bounds for functional tensor trains are provided for Gaussian densities in \cite{Rohrbach:22}. Implementation of point mass filters is inspected in \cite{Matousek:24}, fusion of densities is prospected in \cite{Ajgl:24}. The issue of negative values of density approximations is reported there, but has not been fully explained. Also, several questions such that the grid placement remain to be elaborated.
 
 The goal of this paper is to take a closer look at the behaviour of the approximation process. The presence of areas of negative values is shown to be inherent. The influence of correlation on ranks is demonstrated for different patterns of dependence. The possibility that a decorrelation can have an adverse effect is revealed, even if an interpolation between two grids is replaced by an exact sampling.

 The paper is organised as follows. Section~\ref{sec:formulation} formulates the problem. Section~\ref{sec:analysis} performs a basic analysis. A further discussion is made in Section~\ref{sec:discusion}. The findings are summarised in Section~\ref{sec:summary}.

\section{Problem Formulation}
 \label{sec:formulation}
 Section~\ref{ssec:tt_decomp} introduces both functional and numerical tensor trains. Section~\ref{ssec:mat_decomp} shows the essential matrix decompositions that are used during the construction of tensor trains. Section~\ref{ssec:problem} describes the object of study of this paper.

\newcommand{\stav}{\mathbf{x}}
\newcommand{\sloz}[1]{x_{#1}}
\newcommand{\dimenze}{d}

\newcommand{\hust}{f}
\newcommand{\faktor}[1]{g_{#1}}
\newcommand{\hodn}[1]{r_{#1}}
\newcommand{\indmat}[1]{\alpha_{#1}}

\newcommand{\tenF}{\mathbf{F}}
\newcommand{\tenG}{\mathbf{G}}
\newcommand{\jadro}[1]{\mathbf{G}_{#1}}
\newcommand{\indsloz}[1]{i_{#1}}
\newcommand{\pocsloz}[1]{n_{#1}}
\newcommand{\indj}{j}
\newcommand{\mriz}[1]{\mathbf{x}_{#1}}

 \subsection{Tensor train decompositions}
  \label{ssec:tt_decomp}
  Let $\stav$ be a $\dimenze$-dimensional state with elements $\sloz\indj$, $\indj=1,\ldots,\dimenze$. The functional tensor train decomposition \cite{Gorodetsky:19,Oseledets:13} approximates a scalar-valued function $\hust$ by a product of matrix-valued functions $\faktor\indj$,
  \begin{equation}
   \hust(\stav)=\hust(\sloz1,\ldots,\sloz\dimenze) \approx \faktor1(\sloz1) \ldots \faktor\dimenze(\sloz\dimenze), \label{eq:ftt_matice}
  \end{equation}
  where the first function $\faktor1$ is row-valued and the last function $\faktor\dimenze$ is column-valued. The decomposition can be rewritten by scalar-valued functions $\faktor\indj^{(\indmat{\indj-1},\indmat\indj)}$ for $\indmat\indj=1,\ldots,\hodn\indj$, $\indj=1,\ldots,\dimenze$, where $\hodn\indj$ are called tensor-train ranks and are finite. The approximation \eqref{eq:ftt_matice} can thus be expressed as  
  \begin{equation}
   \hust(\stav) \approx \sum_{\indmat0=1}^{\hodn0} \ldots \sum_{\indmat\dimenze=1}^{\hodn\dimenze} \faktor1^{(\indmat0,\indmat1)}(\sloz1) \ldots \faktor\dimenze^{(\indmat{\dimenze-1},\indmat\dimenze)} (\sloz\dimenze), 
  \end{equation}
  where $\hodn0=\hodn\dimenze=1$ is given by definition.
  
  Tensors can be obtained by sampling the function $\hust$ over grids of points. Let the marginal grids $\mriz\indj$ be given by $\pocsloz\indj$ points, $\mriz\indj=[\mriz\indj(\indsloz\indj)]_{\indsloz\indj=1}^{\pocsloz\indj}$. The sampling given by $\tenF(\indsloz1,\ldots,\indsloz\dimenze) = \hust(\mriz1(\indsloz1),\ldots,\mriz\dimenze(\indsloz\dimenze))$ produces the tensor $\tenF$, i.e.\ a multidimensional array composed of its elements as $\tenF=[\tenF(\indsloz1,\ldots,\indsloz\dimenze)]_{\indsloz1=1,\ldots,\indsloz\dimenze=1}^{\pocsloz1,\ldots,\pocsloz\dimenze}$. The elements can be approximated by a tensor $\tenG$ analogously to \eqref{eq:ftt_matice} as
  \begin{equation}
   \tenF(\indsloz1,\ldots,\indsloz\dimenze) \approx \tenG(\indsloz1,\ldots,\indsloz\dimenze) = \jadro1(\indsloz1) \ldots \jadro\dimenze(\indsloz\dimenze), \label{eq:tt}
  \end{equation}
  where the matrices $\jadro\indj(\indsloz\indj)=\faktor\indj(\mriz\indj(\indsloz\indj))$ have sizes $\hodn{\indj-1} \times \hodn\indj$ and can be organised in three-dimensional arrays $\jadro\indj$ called tensor train cores. The construction of low-rank cores $\jadro\indj$ from the full tensor $\tenF$ can use iterative algorithms or can rely on reshaping of arrays interleaved with matrix decompositions.

 \subsection{Matrix decompositions}
  \label{ssec:mat_decomp}
\newcommand{\matM}{\mathbf{M}}
\newcommand{\matR}{\mathbf{R}}
\newcommand{\matC}{\mathbf{C}}
\newcommand{\matB}{\mathbf{B}}
\newcommand{\matA}{\mathbf{A}}

\newcommand{\hodr}{r}

\newcommand{\indc}{l}
\newcommand{\indr}{k}
\newcommand{\indb}{i}

\newcommand{\matU}{\mathbf{U}}
\newcommand{\matS}{\mathbf{S}}
\newcommand{\matV}{\mathbf{V}}

\newcommand{\T}{\mathrm{T}}
  
  Approximations based on the cross decomposition \cite{Goreinov:97,Savostyanov:14} of a matrix $\matM=[\matM(\indr,\indc)]_{\indr=1,\indc=1}^{\pocsloz\indr,\pocsloz\indc}$ employ the form
  \begin{equation}
   \matM \approx \matA = \matC \matB^{-1} \matR, \label{eq:cross}
  \end{equation}
  where the matrices $\matC$, $\matR$ and $\matB$ are composed from the elements of the matrix $\matM$. Namely, the matrix $\matC$ is composed of $\hodr$ different columns of $\matM$, $\indc_1,\ldots,\indc_\hodr$, the matrix $\matR$ is composed of $\hodr$ different rows of $\matM$, $\indr_1,\ldots,\indr_\hodr$ and the matrix $\matB$ is the submatrix of $\matM$ corresponding to these columns and rows,
  \begin{equation}
   \matC(\indr,\indj)=\matM(\indr,\indc_\indj),\ \matR(\indb,\indc)=\matM(\indr_\indb,\indc),\ \matB(\indb,\indj)=\matM(\indr_\indb,\indc_\indj)
  \end{equation}
  with $\indr=1,\ldots,\pocsloz\indr$, $\indc=1,\ldots,\pocsloz\indc$, $\indb=1,\ldots,\hodr$, $\indj=1,\ldots,\hodr$. If $\hodr$ is equal to the rank of $\matM$, the matrix can be factorised exactly. For lower values of $\hodr$, an approximation is obtained, but the matrix $\matA$ still reconstructs the values of $\matM$ exactly on the columns and rows that have been inspected, i.e.\ it holds $\matM(\indr,\indc_\indj)=\matA(\indr,\indc_\indj)$ and $\matM(\indr_\indb,\indc)=\matA(\indr_\indb,\indc)$. This gave the name to the decomposition. Note also that if an oracle provides the $2 \hodr$ indices $\indr_\indb$ and $\indc_\indj$, no other elements of $\matM$ have to be evaluated in order to compute the decomposition.
  
  The singular value decomposition (SVD) uses the factorisation $\matM = \matU \matS \matV^\T$, where the matrices $\matU$ and $\matV$ are orthonormal and the matrix $\matS$ is rectangular diagonal with non-negative elements on the diagonal sorted in the descending order. The approximation is given by  selecting the first $\hodr$ columns of $\matU$ and $\matV$ and the corresponding submatrix of $\matS$, i.e.\ by
  \begin{equation}
   \matM \approx \matA = \matU_\hodr \matS_\hodr \matV_\hodr^\T, \label{eq:svd}
  \end{equation}
  where $\matU_\hodr(\indr,\indj)=\matU(\indr,\indj)$ for $\indr=1,\ldots,\pocsloz\indr$, $\indj=1,\ldots,\hodr$, $\matV_\hodr(\indc,\indj)=\matV(\indc,\indj)$ for $\indc=1,\ldots,\pocsloz\indc$, $\indj=1,\ldots,\hodr$ and $\matS_\hodr(\indr,\indc)=\matS(\indr,\indc)$ for $\indr=1,\ldots,\hodr$, $\indc=1,\ldots,\hodr$. For a fixed rank $\hodr$, the approximation is optimal in the Frobenius norm, but its downside is that all elements of $\matM$ have to be evaluated.
 
 \subsection{Open issues}
  \label{ssec:problem}
  The first issue is that although all elements of $\matM$ are non-negative, the approximations \eqref{eq:cross}, \eqref{eq:svd} can provide negative values. The second issue is that the order of the factors in \eqref{eq:ftt_matice}, \eqref{eq:tt} is a crucial design parameter and therefore, location of two correlated elements $\sloz\indr$, $\sloz\indc$ influences the memory requirements of the low-rank approximation for a given precision. The third issue is that interpolation between two grids, which is frequently needed in estimation, introduces quantisation noise, which is known to be detrimental to tensor approximations. These issues are studied in the sequel.

\section{Analysis}
 \label{sec:analysis}

 \subsection{Comparison of SVD and cross decompositions}
  \label{ssec:comparison}
  \newcommand{\parm}[1]{\mu_{#1}} 
  \newcommand{\pars}[1]{\sigma_{#1}}
  
  \newcommand{\prva}{a}
  \newcommand{\prvb}{b}

  \begin{figure*}
   \qquad{\includegraphics[scale=0.78]{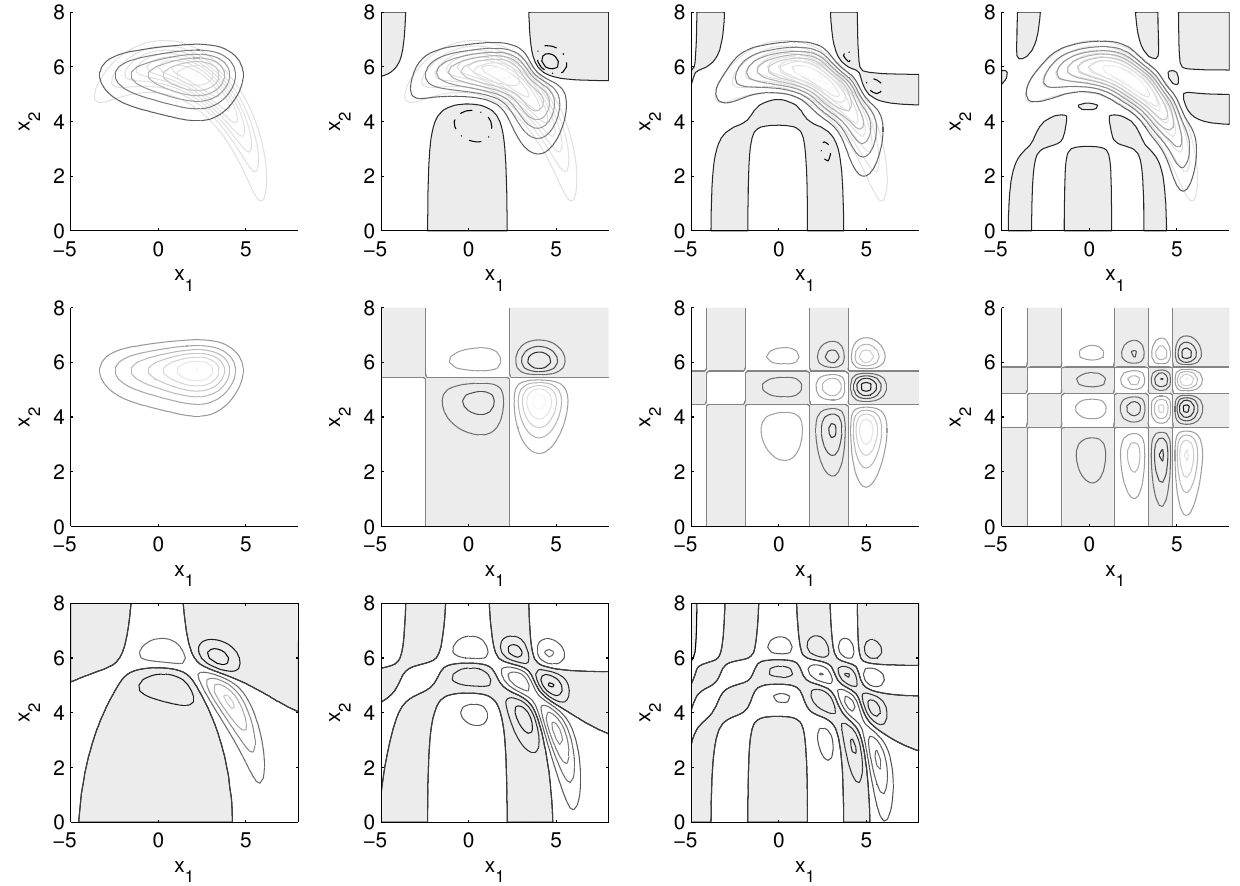}}
   \caption{SVD based approximation: contour lines (greyscale) and areas (shaded) of negative values. The top-row figures show rank-$\hodr$ approximations \eqref{eq:svd} for $\hodr=1,\ldots,4$ and the exact density (light grey contour lines). 
The middle-row figures show the updates, i.e.\ the differences between the consecutive approximations.
The bottom-row figures show approximation errors derived from the top-row figures.}
   \label{fig:svd}
  \end{figure*}
  
  The property of decompositions that has been observed earlier in the literature will be inspected in detail. Namely, aspects of how tensor decompositions approximate tensors of positive density functions by tensors with negative values will be studied on a numerical example.
  
\begin{figure*}
   \centerline{\includegraphics[scale=0.78]{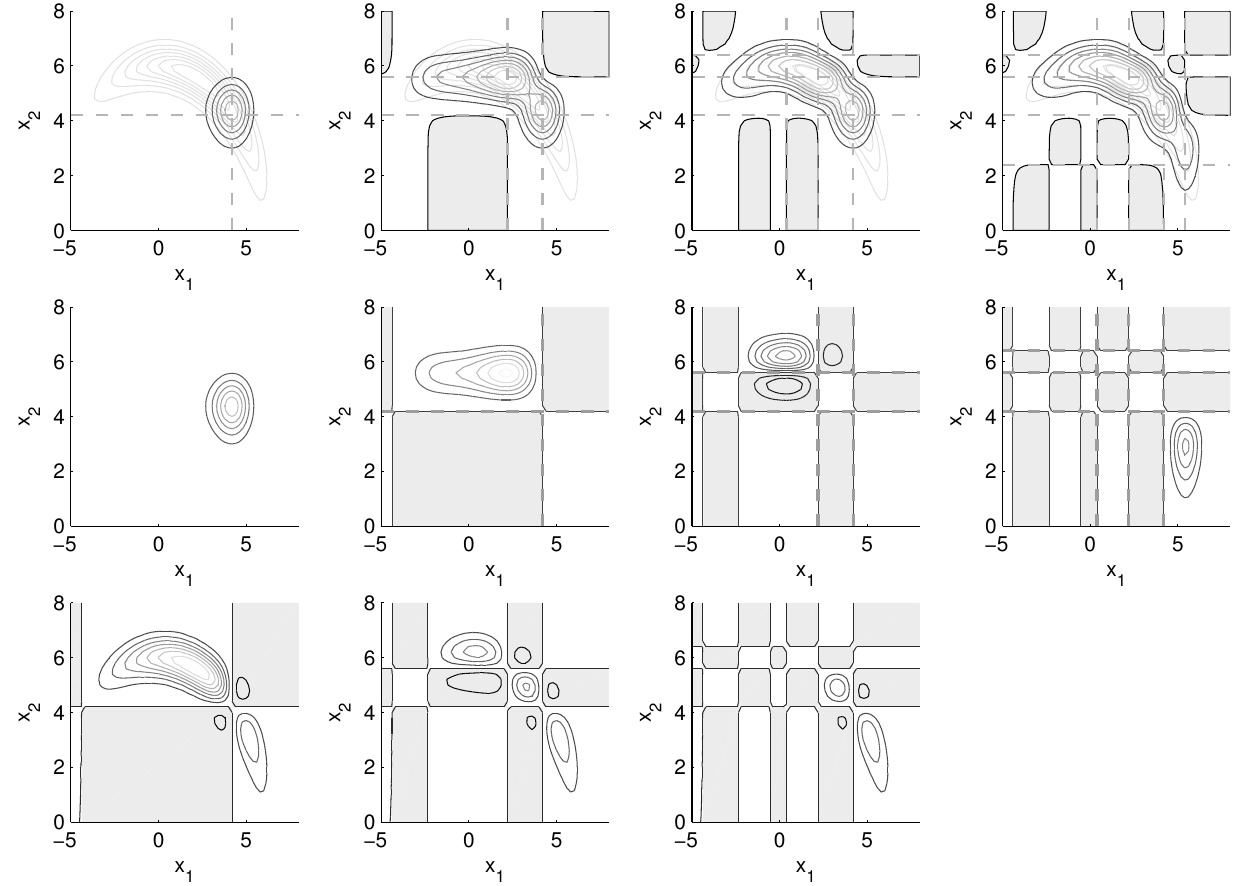}}
   \caption{One run of the cross approximation \eqref{eq:cross}. The meaning corresponds to Fig.~\ref{fig:svd}. The dashed lines in the top-row figures show the grid points with exact representation of the tensor in the top-row figures and zero updates in the middle-row figures.}
   \label{fig:cross}
  \end{figure*}

  Let the state $\stav$ be two-dimensional, $\dimenze=2$. The marginal grids are designed as equidistant as $\mriz1=[-5,-4.8,\ldots,8]$ and $\mriz2=[0,0.2,\ldots,8]$, i.e.\ they have $\pocsloz1=66$ and $\pocsloz2=41$ points. The function $\hust(\stav)$ is chosen as proportional to the truncated Gaussian density for the range and angle from a radar at $[\sloz1,\sloz2]=[0,0]$ with means and standard deviations given by $\parm{r}=6$, $\pars{r}=0.5$ and $\parm{a}=1.2$, $\pars{a}=0.5$, respectively, but shown in Cartesian coordinates, $\hust(\sloz1,\sloz2)=(2 \pi \pars{r} \pars{a})^{-1}*$ $\exp\big(-\frac{1}{2}(\sloz1^2+\sloz2^2-\parm{r})^2 \pars{r}^{-2}-\frac{1}{2}(\operatorname{atan2}(\sloz2,\sloz1)-\parm{a})^2 \pars{a}^{-2}\big)$. The approximations are illustrated in Figs.~\ref{fig:svd} and~\ref{fig:cross} for SVD and cross decompositions, respectively.

  Fig.~\ref{fig:svd} shows the SVD decomposition. The first four singular values are $5.78$, $3.14$, $1.80$ and $1.00$ and the corresponding approximations \eqref{eq:svd} for $\hodr=1,\ldots,4$ are shown in the top-row figures. It can be observed how the approximations converge to the density and that areas of negative values appear and evolve during this process. The explanation of the occurrence can be inferred from the other rows. The bottom figures show the error of the approximations shown in the top figures. These functions are approximated by rank-$1$ functions shown in the middle figures, while the panels are shifted one position to the right. That is, the updates of the rank-$1$ approximation to rank-$2$, rank-$3$ and rank-$4$ approximations are shown in the second, third and fourth middle-row figures. The functions are products of the corresponding columns of $\matU$ and $\matV$, which are orthogonal to the first columns of $\matU$ and $\matV$. The first columns are either non-negative or non-positive by the construction of rank-$1$ approximation of a non-negative function. Therefore, except the first columns of $\matU$ and $\matV$, all other columns have to switch the sign. The updates inevitably contain areas of negative values, which introduces such areas to the density approximations. The amplitude of the updates decreases with the singular values, the negative values in the approximations approach zero and the shaded areas in the top figures would disappear for the full-rank decomposition and infinite precision of numbers.

  Fig.~\ref{fig:cross} shows the cross decomposition. The layout of the figures is the same as in Fig.~\ref{fig:svd}. The cross decomposition is not recursive in the sense that the best rank-$\hodr$ approximation does not update the best rank-$(\hodr-1)$ approximation. That is, the best selections $\{\indc_1,\ldots,\indc_\hodr\}$ and $\{\indr_1,\ldots,\indr_\hodr\}$ for fixed $\hodr$ are not subsets of the best selections for $\hodr+1$.  Efficient suboptimal algorithms are stochastic and search the approximations heuristically. One run of the \texttt{greedy2\_cross} algorithm \cite{Savostyanov:14} is illustrated. The points $\mriz1(\indc_\indj)$ and $\mriz2(\indr_\indj)$, $\indj=1,\ldots,4$, are given by $\{4.2, 2.2, 0.4, 5.4\}$ and $\{4.2, 5.6, 6.4, 2.4\}$, respectively. Now, the approximations are numerically exact on the grid points in the selected columns and rows, which are shown by the dashed lines in the top figures. Therefore, the following updates have to be zero there as well, which is indicated in the middle figures and correspond to some zero-level contours. The introduction of negative values in \eqref{eq:cross} for $\matM$ with all values being positive can be best illustrated for $\matB=\begin{bsmallmatrix}\prva & \prvb\\ \prvb & \prva \end{bsmallmatrix}$ with $\prva>\prvb>0$. Since the inverse $\matB^{-1}$ is $\frac{1}{\prva^2-\prvb^2}\begin{bsmallmatrix}\prva & -\prvb\\ -\prvb & \prva \end{bsmallmatrix}$, it introduces differences of two rows or two column into the decomposition. Again, the areas of positive and negative updates form a chequered pattern. Contrary to the SVD, the cross decomposition seems to non-negligibly update the approximation of density functions in small areas only. It is therefore considerably slower. Finally, it has to be noted that for low ranks~$\hodr$, many of other runs of the algorithm have provided worse approximations than the results shown above.
  
  In summary, both SVD and cross decompositions of matrices provide approximations with negative values. The SVD decomposition shrinks the approximation errors over whole grid, the cross decomposition fixes the values at an increasing number of grid point. The updates have chequered patterns, which creates complex areas of negative values. With an increasing rank, the minimal values tend to approach zero, but the size of the areas does not approach zero for low ranks.
  
  The takeaway is that for these tensor decompositions of multidimensional arrays, low-rank approximations of densities will contain negative values at many grid points. This precludes practicability of some concepts. Namely, the idea of approximating the square root of a probability density function and squaring it in order to obtain the original density is not feasible. The reason is that the squaring will boost the ranks and therefore, a low-rank approximation of the square will be taken for the tensor rounding. Since the same techniques are used for the rounding as for the original approximation, the rounding will bring back the areas of negative values.

  Last, note that direct non-negative decompositions exist, see e.g.\ \cite{Bhattarai:20} and the references therein. However, these methods are not cheap and lack adaptivity. They are based on optimisation procedures and require the tensor-train ranks to be preselected, which means that it is not straightforward to design an approximation with a prescribed precision. Moreover, the ranks of the non-negative decompositions are higher than the ranks of the unconstrained decomposition for the same precision level. A new non-negative rounding procedure for tensor trains would be needed, since the procedure is used intensively in filtering.

  \newcommand{\kov}[1][]{\mathbf{Q}_{#1}}
  \newcommand{\cast}[1]{\stav_{#1}}
  \newcommand{\zacatek}{a}
  \newcommand{\stred}{b}
  \newcommand{\konec}{c}
  \newcommand{\kandidat}[1]{\bar{g}_{#1}}
  \newcommand{\korekce}{\mathbf{D}}
  \newcommand{\matId}[1][]{\mathbf{I}_{#1}}
  \newcommand{\matnul}{\mathbf{0}}
  \newcommand{\indz}{\alpha}
  \newcommand{\inds}{\beta}
  \newcommand{\indk}{\gamma}
  \newcommand{\ming}{\delta}
  \begin{figure*}
   \begin{align}
    &\stav^\T \kov \stav = 
     \begin{bmatrix}\cast\zacatek \\ \cast\stred \\ \cast\konec \end{bmatrix}^T
     \begin{bmatrix}\kov[\zacatek,\zacatek] & \kov[\zacatek,\stred] & \kov[\zacatek,\konec] \\ \kov[\stred,\zacatek] & \kov[\stred,\stred] & \kov[\stred,\konec] \\  \kov[\konec,\zacatek] & \kov[\konec,\stred] & \kov[\konec,\konec] \end{bmatrix}
     \begin{bmatrix}\cast\zacatek \\ \cast\stred \\ \cast\konec \end{bmatrix}
    =\begin{bmatrix} 1 \\ \cast\zacatek \\ \cast\zacatek^\T \kov[\zacatek,\zacatek] \cast\zacatek \end{bmatrix}^\T
     \begin{bmatrix}1 & \cast\stred^\T (\kov[\stred,\konec]+\kov[\konec,\stred]^\T) & \cast\stred^T \kov[\stred,\stred] \cast\stred \\ \matnul & \kov[\zacatek,\konec]+\kov[\konec,\zacatek]^\T & (\kov[\zacatek,\stred] + \kov[\stred,\zacatek]^\T) \cast\stred \\ 0 & \matnul & 1\end{bmatrix}
     \begin{bmatrix}\cast\konec^\T \kov[\konec,\konec] \cast\konec  \\ \cast\konec \\ 1\end{bmatrix} \label{eq:kv_fun_rozklad}
   \end{align}
  \end{figure*}

 \subsection{Quadratic function and Gaussian function}
  \label{ssec:quadratic_fun}

  The analysis continues with higher dimensions. A multivariate quadratic function is discussed first. A Gaussian function, which is proportional to the exponential of a quadratic function, is inspected next for the four-variate case. Namely, the relation of functional tensor train decomposition of a quadratic function with tensor train decomposition of sampled Gaussian densities is stressed.
  
  A quadratic function $\hust$ of a $\dimenze$-dimensional state $\stav$ is given by a square matrix $\kov$ of the size $\dimenze$ as $\hust(\stav)=\stav^\T \kov \stav$. If the state is partitioned into three sub-vectors $\cast\zacatek$, $\cast\stred$ and $\cast\konec$ and the matrix $\kov$ into corresponding nine blocks, the basic algebra provides the factorisation \eqref{eq:kv_fun_rozklad}, where each factor is a matrix-valued function of the individual vector arguments. Without a loss of generality, it can be assumed that the matrix $\kov$ is symmetric. Thus, only the upper triangular elements may be considered.

  Exact functional tensor trains \eqref{eq:ftt_matice} of a quadratic function can be constructed \cite{Tichavsky:24,Gorodetsky:19}. Many parametrisations are possible. Based on \eqref{eq:kv_fun_rozklad}, the following matrix functions are designed,
  \begin{align}
   &\kandidat\indj(\sloz\indj)=\nonumber\\
   &\left\{\begin{array}{l l}
   \begin{bmatrix}1 & \sloz1 & \kov[1,1] \sloz1^2 \end{bmatrix}& \indj=1\\[2pt]
   \begin{bmatrix}1 & \matnul & \sloz\indj & \kov[\indj,\indj] \sloz\indj^2\\ \matnul & \matId[\indj-1] & \matnul & 2 \kov[1:\indj-1,\indj] \sloz\indj \\ 0 &\matnul &0 & 1\end{bmatrix} & 1<\indj<\frac{d+1}{2}\\[2pt]
   \begin{bmatrix}1 &0 &\matnul &0\\ 2 \kov[\indj,\indj+1:\dimenze]^\T \sloz\indj & \matnul & \matId[\dimenze-\indj] & \matnul \\ \kov[\indj,\indj] \sloz\indj^2 & \sloz\indj & \matnul & 1\end{bmatrix}^\T & \frac{d-1}{2}<\indj<\dimenze \\[2pt]
   \begin{bmatrix}\kov[\dimenze,\dimenze] \sloz\dimenze^2 & \sloz\dimenze & 1\end{bmatrix}^\T & \indj=\dimenze
   \end{array}\right.. \label{eq:kand_j}
  \end{align}
  For an index $\indz<\frac{\dimenze+1}{2}$, the product $\kandidat1(\sloz1)\ldots\kandidat\indz(\sloz\indz)$ leads to the left factor in \eqref{eq:kv_fun_rozklad}, i.e.\ $[1,\cast\zacatek^\T,\cast\zacatek^\T \kov[\zacatek,\zacatek]\cast\zacatek]$ with $\cast\zacatek=[\sloz1,\ldots,\sloz\indz]^\T$ and the corresponding block $\kov[\zacatek,\zacatek]$. Similarly, an index $\indk>\frac{\dimenze+1}{2}$ and product $\kandidat\indk(\sloz\indk)\ldots\kandidat\dimenze(\sloz\dimenze)$ lead to the right factor in \eqref{eq:kv_fun_rozklad}, i.e.\ $[\cast\konec^\T \kov[\konec,\konec] \cast\konec,\cast\konec^\T, 1]^T$ with $\cast\konec=[\sloz\indk,\ldots,\sloz\dimenze]^\T$. 
  
  For odd dimension $\dimenze$, the remaining function for $\indj=\frac{\dimenze+1}{2}$ is designed as 
  \begin{equation}
   \kandidat\indj(\sloz\indj)|_{\indj=\frac{\dimenze+1}{2}} = \begin{bmatrix}1 & 2 \sloz{\indj} \kov[\indj,\indj+1:\dimenze] & \kov[\indj,\indj] \sloz{\indj}^2 \\ \matnul & 2 \kov[1:\indj-1,\indj+1:\dimenze] & 2 \kov[1:\indj-1,\indj] \sloz{\indj} \\ 0 & \matnul & 1 \end{bmatrix}
  \end{equation}
  and corresponds to the middle factor in \eqref{eq:kv_fun_rozklad} for $\cast\stred=\sloz{(\dimenze+1)/2}$. The cores $\faktor\indj(\sloz\indj)$ of the functional tensor train \eqref{eq:ftt_matice} are given by the candidate functions $\kandidat\indj(\sloz\indj)$ directly for all $\indj=1,\ldots,\dimenze$.

  For even dimension $\dimenze$, the middle factor in \eqref{eq:kv_fun_rozklad} corresponds to an empty part $\cast\stred$. A correction matrix $\korekce$ is designed as 
  \begin{equation}
   \korekce\overset{\dimenze \text{ even}}{=}\begin{bmatrix}1 & \matnul & 0 \\ \matnul & 2 \kov[1:\frac{\dimenze}{2},\frac{\dimenze}{2}+1:\dimenze] & \matnul\\ 0 & \matnul & 1 \end{bmatrix} \label{eq:kand_kor}
  \end{equation}
  and has to be incorporated into one of the neighbouring factors. That it, it holds $\faktor\indj(\sloz\indj)=\kandidat\indj(\sloz\indj)$ for all $\indj$ except $\indj=\frac{\dimenze}{2}$, where $\faktor{\frac{\dimenze}{2}}(\sloz{\frac{\dimenze}{2}})=\kandidat{\frac{\dimenze}{2}}(\sloz{\frac{\dimenze}{2}}) \korekce$ holds, or alternatively for all $\indj$ except $\indj=\frac{\dimenze}{2}+1$ and $\faktor{\frac{\dimenze}{2}+1}(\sloz{\frac{\dimenze}{2}}+1)=\korekce \kandidat{\frac{\dimenze}{2}+1}(\sloz{\frac{\dimenze}{2}}+1)$.
    
  Table~\ref{tab:kvadr_fun} shows the cores $\faktor\indj(\sloz\indj)$ for $\dimenze=4$. All diagonal elements $\kov[\indj,\indj]$ are considered non-zero. Various sets of non-zero upper triangular elements $\kov[\indr,\indc]$ are inspected (with lower triangular elements $\kov[\indc,\indr]=\kov[\indr,\indc]$ by definition) and denoted by the pairs $(\indr,\indc)$. It is easy to observe that for several zero off-diagonal elements, some rows or columns in \eqref{eq:kand_j}, \eqref{eq:kand_kor} become zero. In such cases, some rows and columns can be eliminated. The squeezed cores are shown in the table and the corresponding tensor train ranks are shown in Table~\ref{tab:kvadr_hod}.
  \begin{table*}[t]
   \caption{Cores of functional tensor trains for a quadratic function $\stav^\T \kov \stav$ for cases $(\indr,\indc)$ of non-zero upper triangular elements $\kov[\indr,\indc]$.}
   \begin{center}
    $
    \begin{array}{|c||c|c|c|c|}
     \hline
     \text{case} & \faktor1(\sloz1) & \faktor2(\sloz2) & \faktor3(\sloz3) & \faktor4(\sloz4) \\
     \hline\hline
     \emptyset & \begin{bmatrix}1 & \kov[1,1] \sloz1^2\end{bmatrix} & \begin{bmatrix}1 & \kov[2,2] \sloz2^2 \\ 0 & 1\end{bmatrix}& \begin{bmatrix}1 & \kov[3,3] \sloz3^2 \\ 0 & 1\end{bmatrix} & \begin{bmatrix}\kov[4,4] \sloz4^2 \\ 1\end{bmatrix}\\
     \hline
     (2,3) & \begin{bmatrix}1 & \kov[1,1] \sloz1^2\end{bmatrix} & \begin{bmatrix}1 & 2 \kov[2,3] \sloz2 & \kov[2,2] \sloz2^2 \\ 0 & 0 & 1\end{bmatrix}& \begin{bmatrix}1 & \kov[3,3] \sloz3^2 \\ 0 &\sloz3\\ 0 & 1\end{bmatrix} & \begin{bmatrix}\kov[4,4] \sloz4^2 \\ 1\end{bmatrix}\\
     \hline
     (1,2) & \begin{bmatrix}1 & \sloz1 & \kov[1,1] \sloz1^2\end{bmatrix} & \begin{bmatrix}1 & \kov[2,2] \sloz2^2 \\ 0 & 2 \kov[1,2] \sloz2 \\ 0 & 1\end{bmatrix}& \begin{bmatrix}1 & \kov[3,3] \sloz3^2 \\ 0 & 1\end{bmatrix} & \begin{bmatrix}\kov[4,4] \sloz4^2 \\ 1\end{bmatrix}\\
     \hline
     (1,3) & \begin{bmatrix}1 & \sloz1 & \kov[1,1] \sloz1^2\end{bmatrix} & \begin{bmatrix}1 & 0 & \kov[2,2] \sloz2^2 \\ 0 & 2 \kov[1,3] & 0 \\ 0 & 0 & 1\end{bmatrix}& \begin{bmatrix}1 & \kov[3,3] \sloz3^2 \\ 0 & \sloz3\\ 0 & 1\end{bmatrix} & \begin{bmatrix}\kov[4,4] \sloz4^2 \\ 1\end{bmatrix}\\
     \hline
     (1,4) & \begin{bmatrix}1 & \sloz1 & \kov[1,1] \sloz1^2\end{bmatrix} & \begin{bmatrix}1 & 0 & \kov[2,2] \sloz2^2 \\ 0 & 2 \kov[1,4] & 0 \\ 0 & 0 & 1\end{bmatrix}& \begin{bmatrix}1 & 0 & \kov[3,3] \sloz3^2 \\ 0 & 1 & 0\\ 0 & 0 & 1\end{bmatrix} & \begin{bmatrix}\kov[4,4] \sloz4^2 \\ \sloz4 \\ 1\end{bmatrix}\\
     \hline
     (1,2), (3,4) & \begin{bmatrix}1 & \sloz1 & \kov[1,1] \sloz1^2\end{bmatrix} & \begin{bmatrix}1 & \kov[2,2] \sloz2^2 \\ 0 & 2 \kov[1,2] \sloz2 \\ 0 & 1\end{bmatrix}& \begin{bmatrix}1 & 2 \kov[3,4] \sloz3 & \kov[3,3] \sloz3^2 \\ 0 & 0 & 1\end{bmatrix} & \begin{bmatrix}\kov[4,4] \sloz4^2 \\ \sloz4 \\ 1\end{bmatrix}\\
     \hline
     (1,3), (2,4) & \begin{bmatrix}1 & \sloz1 & \kov[1,1] \sloz1^2\end{bmatrix} & \begin{bmatrix}1 & 0 & 2 \sloz2 \kov[2,4] & \kov[2,2] \sloz2^2 \\ 0 & 2 \kov[1,3] & 0 & 0 \\ 0 & 0 & 0 & 1\end{bmatrix}& \begin{bmatrix}1 & 0 & \kov[3,3] \sloz3^2 \\ 0 & 0 &\sloz3\\ 0 & 1 & 0\\ 0 & 0 & 1\end{bmatrix} & \begin{bmatrix}\kov[4,4] \sloz4^2 \\ \sloz4 \\ 1\end{bmatrix}\\
     \hline
     (1,4), (2,3) & \begin{bmatrix}1 & \sloz1 & \kov[1,1] \sloz1^2\end{bmatrix} & \begin{bmatrix}1 & 2 \kov[2,3] \sloz2 & 0 & \kov[2,2] \sloz2^2 \\ 0 & 0 & 2 \kov[1,4] & 0 \\ 0 & 0 & 0 & 1\end{bmatrix}& \begin{bmatrix}1 & 0 & \kov[3,3] \sloz3^2 \\ 0 & 0 &\sloz3\\ 0 & 1 & 0\\ 0 & 0 & 1\end{bmatrix} & \begin{bmatrix}\kov[4,4] \sloz4^2 \\ \sloz4 \\ 1\end{bmatrix}\\
     \hline
     \text{full} & \begin{bmatrix}1 & \sloz1 & \kov[1,1] \sloz1^2\end{bmatrix} & \begin{bmatrix}1 & 2 \kov[2,3] \sloz2 & 2 \sloz2 \kov[2,4] & \kov[2,2] \sloz2^2 \\ 0 & 2 \kov[1,3] & 2 \kov[1,4] & 2 \kov[1,2] \sloz2 \\ 0 & 0 & 0 & 1\end{bmatrix}& \begin{bmatrix}1 & 2\kov[3,4] \sloz3 & \kov[3,3] \sloz3^2 \\ 0 & 0 &\sloz3\\ 0 & 1 & 0\\ 0 & 0 & 1\end{bmatrix} & \begin{bmatrix}\kov[4,4] \sloz4^2 \\ \sloz4 \\ 1\end{bmatrix}\\
     \hline
    \end{array}
    $
    \label{tab:kvadr_fun}
   \end{center}
  \end{table*}
  
  The case of a diagonal matrix $\kov$ corresponds to the sum of functions of the individual elements $\sloz\indj$ and therefore, all ranks $\hodn1,\ldots,\hodn{\dimenze-1}$ are equal to $2$, see e.g.\ \cite{Gorodetsky:19,Oseledets:13}. The following observations are made. If a single non-zero upper triangular element $\kov[\indj,\indj+1]$ is considered for neighbouring elements $\sloz\indj$ and $\sloz{\indj+1}$, a single rank $\hodn\indj$ is increased to $3$. If a single non-zero element $\kov[\indr,\indc]$ is considered for non-neighbouring elements $\sloz\indr$ and $\sloz\indc$, the ranks $\hodn\indr,\ldots,\hodn{\indc-1}$ are increased to $3$. If two elements $\kov[\indb,\indj]$ and $\kov[\indr,\indc]$ with $\indb<\indj<\indc<\indr$ are considered, the ranks $\hodn\indb,\ldots,\hodn{\indj-1}$ and $\hodn\indr,\ldots,\hodn{\indc-1}$ are increased to $3$. In the case $\indb<\indc<\indj<\indr$, the ranks $\hodn\indb,\ldots,\hodn{\indc-1}$ and $\hodn\indj,\ldots,\hodn{\indr-1}$ are increased to $3$ and the ranks $\hodn\indc,\ldots,\hodn{\indj-1}$ to $4$. The case $\indb<\indc<\indr<\indj$ leads to $\hodn\indb,\ldots,\hodn{\indc-1}$ and $\hodn\indr,\ldots,\hodn{\indj-1}$ being $3$ and $\hodn\indc,\ldots,\hodn{\indr-1}$ being $4$. If all elements of $\kov$ are non-zero, the ranks $\hodn\indj$ increase from $\hodn1=3$ by one for $\indj\le\frac{\dimenze}{2}$  and decrease symmetrically to $\hodn{\dimenze-1}=3$.
  
  Next, experiments with Gaussian densities are performed. Sampling over grids of points is considered. The marginal grids are designed as equidistant as $\mriz\indj=[-4,-3.8,\ldots,4]$, i.e.\ they have $\pocsloz\indj=41$ points. The full grid has $41^4=2825761$ points. Gaussian densities with zero mean and with covariance matrix given by $\kov$ from the preceding examples are considered. All diagonal elements are chosen as unit, $\kov[\indj,\indj]=1$, those off-diagonal elements that are non-zero are chosen as equal to $0.5$. The off-diagonal elements are correlation coefficients. Tensor trains $\tenG$ are computed by the basic SVD method implemented in the \texttt{tt\_tensor} function \cite{Oseledets:TTT} with the precision parameter set as $10^{-5}$. Contrary to the functional trains, the approximations are not exact and the tensor train ranks depend heavily on the designed parameters.

  \begin{table}
   \caption{Ranks of the cores from Table~\ref{tab:kvadr_fun} for $(\indr,\indc)$ cases. }
   \begin{center}
    \renewcommand*{\arraystretch}{1.03}
    $
    \begin{array}{|c||c|c|c|c|c|}
     \hline
     \text{cases of }\kov[\indr,\indc]\neq 0 & \hodn0 & \hodn1 & \hodn2 & \hodn3 & \hodn4\\
     \hline\hline
     \emptyset & 1 & 2 & 2 & 2 & 1\\
     \hline
     (2,3) & 1 & 2 & 3 & 2 & 1\\
     \hline
     (1,2) & 1 & 3 & 2 & 2 & 1\\
     \hline
     (1,3) & 1 & 3 & 3 & 2 & 1\\
     \hline
     (1,4) & 1 & 3 & 3 & 3 & 1\\
     \hline
     (1,2), (3,4) & 1 & 3 & 2 & 3 & 1\\
     \hline
     (1,3), (2,4) & 1 & 3 & 4 & 3 & 1\\
     \hline
     (1,4), (2,3) & 1 & 3 & 4 & 3 & 1\\
     \hline
     \text{full} & 1 & 3 & 4 & 3 & 1\\
     \hline
    \end{array}
    $
    \label{tab:kvadr_hod}
   \end{center}
  \end{table}

  Table~\ref{tab:Gauss_hod} shows the resulting ranks, the pattern is the same as in Table~\ref{tab:kvadr_hod}.  
Further, the numbers of negative values of the approximate tensors $\tenG$ over all $41^4$ grid points are counted. For independent components, there is no negative value. For a pair of correlated components, the considered density with the designed grid lead to $6.42\%$ negative values and this ratio does not depend substantially on the choice of the pair. For two pairs, the ratio rises to the order of tens of percent and it depends on the choice of the correlated pairs. This supports the idea described earlier, i.e.\ that the approximations of density can be negative at a large number of grid points. 

  \begin{table}
   \caption{Ranks of the cores $\jadro\indj$ for non-zero correlation coefficients at $(\indr,\indc)$ given by $0.5$ and precision parameter $10^{-5}$ of \texttt{tt\_tensor}. The number of negative values of $\tenG$.}
   \begin{center}
    \renewcommand*{\arraystretch}{1.05}
    $
    \begin{array}{|c||c|c|c|c|c||c|c|}
     \hline
     \text{cases of }\kov[\indr,\indc]\neq 0 & \hodn0 & \hodn1 & \hodn2 & \hodn3 & \hodn4 & \tenG<0\\
     \hline\hline
     \emptyset & 1 & 1 & 1 & 1 & 1 & 0\%\\
     \hline
     (2,3) & 1 & 1 & 10 & 1 & 1 & 6.42\%\\
     \hline
     (1,2) & 1 & 10 & 1 & 1 & 1 & 6.42\%\\
     \hline
     (1,3) & 1 & 10 & 10 & 1 & 1 & 6.42\%\\
     \hline
     (1,4) & 1 & 10 & 10 & 10 & 1 & 6.42\%\\
     \hline
     (1,2), (3,4) & 1 & 10 & 1 & 10 & 1 & 12.02\%\\
     \hline
     (1,3), (2,4) & 1 & 10 & 55 & 10 & 1 & 16.59\%\\
     \hline
     (1,4), (2,3) & 1 & 10 & 55 & 10 & 1 & 16.59\%\\
     \hline
    \end{array}
    $
    \label{tab:Gauss_hod}
   \end{center}
  \end{table}

  Alternative constructions that do not strictly adhere the precision parameter are described in the sequel. For independent components $\sloz\indj$, all ranks are equal to $1$ and the tensor train is exact. The cores $\jadro\indj$ are constructed as the marginal Gaussian densities, $\jadro\indj(\indsloz\indj)=\hust_\indj\big(\mriz\indj(\indsloz\indj))$. If merely two components $\sloz\indr$ and $\sloz\indc$ are correlated, the ranks $\hodn\indr,\ldots,\hodn{\indc-1}$ increase to the same value $\hodr$. The cores $\jadro\indr$ and $\jadro\indc$ can be constructed by the basic two-dimensional matrix decomposition from Section~\ref{ssec:comparison}, while the cores $\jadro\indj$, $\indj=\indr+1,\ldots,\indc-1$, can be constructed as identity matrices of the size $\hodr$ scaled by the values of the marginal density $\hust_\indj$, $\jadro\indj(\indsloz\indj)=\matId[\hodr] \hust_\indj\big(\mriz\indj(\indsloz\indj)\big)$, and the remaining cores by the values of the remaining marginal densities directly, since the remaining ranks are equal to one. For two correlated pairs $(\sloz\indb,\sloz\indj)$, $(\sloz\indr,\sloz\indc)$ such that $\indb<\indj<\indc<\indr$ holds, two two-dimensional matrix decompositions can be combined and the cores between $\indb$ and $\indj$ and between $\indc$ and $\indr$ handled analogously to the previous case, see the case $(1,2), (3,4)$ in the table. An alternative construction is that two four-dimensional tensor trains for the marginal  densities $\hust_{1,2}(\sloz1,\sloz2)$ and $\hust_{3,4}(\sloz3,\sloz4)$ are constructed and multiplied element-wise according to \cite{Oseledets:11}, i.e.\ the ranks $[1,10,1,1,1]$ and $[1,1,1,10,1]$ are multiplied element-wise to $[1,10,1,10,1]$. Using this approach, the cases $(1,3), (2,4)$ and $(1,4), (2,3)$ can produce the ranks $[1,10,100,10,1]$ either from the product of $[1,10,10,1,1]$ and $[1,1,10,10,1]$ or the product of $[1,10,10,10,1]$ and $[1,1,10,1,1]$. However, using of the rounding procedure \cite{Oseledets:11} for a prescribed precision parameter can reduce the ranks to $[1,10,55,10,1]$.

\section{Discussion}
 \label{sec:discusion}
  \newcommand{\ystav}{\mathbf{y}}
  \newcommand{\ymriz}[1]{\mathbf{y}_{#1}}
  \newcommand{\ykov}{\mathbf{P}}
  \newcommand{\yodm}{\mathbf{R}}
  \newcommand{\strvek}{\mathbf{m}}

  The issue of tensor train ranks is revisited in this section. The question of grid placement and decorrelation is inspected. The numerical example considers the case introduced in Section~\ref{ssec:comparison}.

  In the case of Gaussian densities, higher correlation coefficients lead to higher tensor train ranks. The recommended solution is to design the grids in new coordinates, such that independent components are obtained, which leads to unit ranks. For general densities, the components can be merely decorrelated, i.e.\ the ranks need not be reduced, which is shown in the sequel.

  Let $\strvek$ and $\kov$ be the mean vector and covariance matrix corresponding to the the density $\hust(\stav)$ represented over grid of points with marginal grids $\mriz\indj$. The new coordinates $\ystav$ are introduced by $\ystav=\yodm^{-1} (\stav-\strvek)$, where the matrix $\yodm$ is an arbitrary square root term of $\kov$ given by $\kov = \yodm \yodm^\T$. Then, the components of $\ystav$ become decorrelated.
  
\begin{figure}
   \centerline{\includegraphics[scale=0.8]{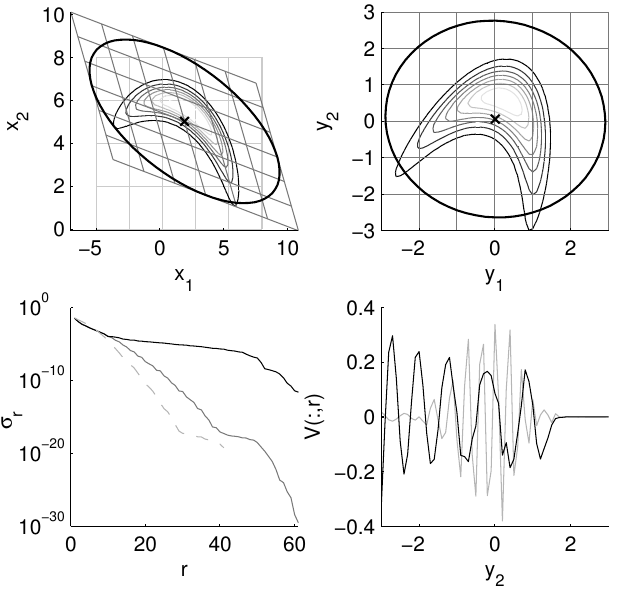}}
   \caption{Interpolation between grids. Top left figure: the function (contour lines), $3$-$\sigma$ ellipse (black) and two grids (light and dark grey). Top right figure: interpolation into the new grid (dark grey). Lower left figure: singular value corresponding to the function over the original grid (dashed line), function over the new grid (dark grey line) and interpolated function over the new grid (black line). Lower right figure: selected singular vectors (black, grey).}
   \label{fig:grid}
  \end{figure}

  Fig.~\ref{fig:grid} shows the normalised function from Fig.~\ref{fig:svd}. The top left figure compares the original grid (grey lines -- with only every thirteenth point in $\mriz1$ and every tenth point in $\mriz2$ drawn for visual clarity) with the new grid (black lines -- with only every tenth point in each $\ymriz\indj$ drawn for visual clarity) for $\ymriz1=\ymriz2=[-3,-2.9,\ldots,3]$ and the symmetric square root of $\kov$, i.e.\ $\yodm=\yodm^\T$. Next, the $3$-$\sigma$ ellipse is shown. The normalised transformed function is shown in the top right figure, where a new $3$-$\sigma$ ellipse is shown. It has to be noted that the mean value is not exactly in the origin $[0,0]$ and that the new ellipse does not exactly match the new grid. The reason is that the numerical integration is not exact. The singular values are shown in the bottom left figure, the dashed line corresponds to $\stav$ coordinates, the grey line corresponds to the $\ystav$. It can be observed that in this particular case, the decorrelation leads to a slower decrease of the singular values. To achieve the same precision of the tensor train approximation, the tensor train ranks have to be higher in the decorrelated coordinates in this example.

  Next, Fig.~\ref{fig:grid} shows the consequences of interpolation. In practice, the functional description is not available in the point mass filter and the values at the new grid have to be constructed based on the values at the old grid. The contour lines at the top right figures have actually been obtained by the linear interpolation for the points within the range of the old grid and taking the value at the nearest point of the old grid for the new points outside the range of the old grid. The singular values of the interpolation are shown in the bottom left figure by the black line. It can be observed that the first eight singular values roughly correspond to the singular values of the exact sampling of the density, compare the black and grey lines. The further singular values decrease very slowly and the higher-rank approximations merely track the interpolation-introduced noise. This can also be observed from the singular vectors. The bottom right figure shows the tenth and eleventh column of the matrix $\matV$ (black and grey lines, respectively). Although the black line keeps the marginal structure known from Fig.~\ref{fig:svd}, the grey line becomes very noisy. Thus, the interpolation to a new grid disables the approximation to be precise, which has to be taken in mind when tensor trains are constructed.

\begin{figure}
   \centerline{\includegraphics[scale=0.8]{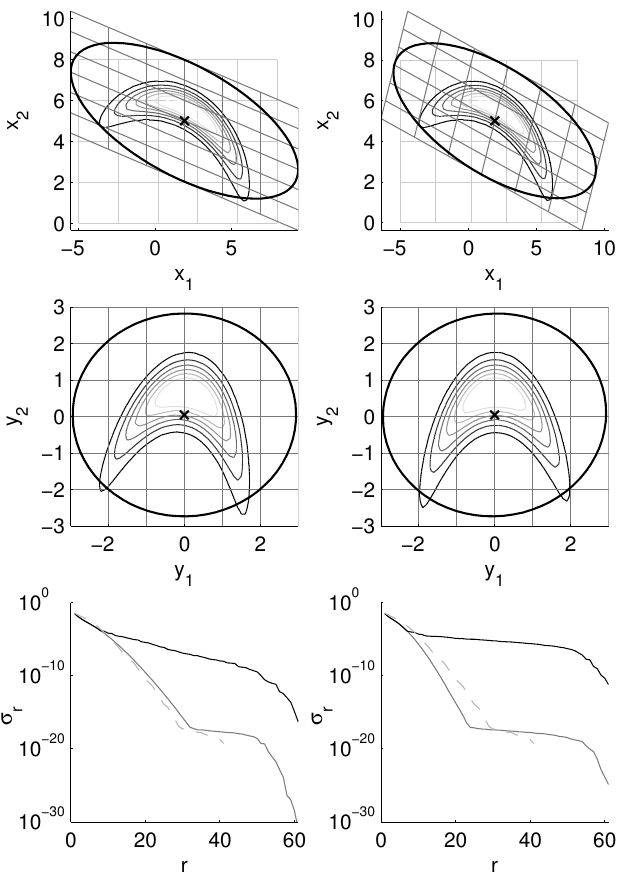}}
   \caption{Other choices of square roots $\yodm$ of $\kov$. The meaning corresponds to Fig.~\ref{fig:grid}. Left: Cholesky decomposition. Right: eigenvalue-based decomposition.}
   \label{fig:grot}
  \end{figure}

  It has to be stressed that the results depend on the choice of the square root $\yodm$ heavily. Fig.~\ref{fig:grot} repeats the experiment for two other square roots. The left figures show the Cholesky decomposition, i.e.\ $\yodm$ is a lower-triangular matrix. The right figures show the eigenvalue-based decompositions, i.e.\ the columns of $\yodm$ are given by left eigenvectors scaled by the square roots of the corresponding eigenvalues of $\kov$.
  The decorrelation of the normalised function can achieve better rates than before. For example, the eigenvalue-based square root is the best in terms of the exact transformation. However, it is the worst if the interpolation\slash extrapolation is incorporated, i.e. the black line at the bottom right figure shows the slowest decrease out of the three considered cases of $\yodm$. Unfortunately, there is no guideline on the design of the best square root.

\section{Summary}
 \label{sec:summary}
 Application of tensor train decompositions for approximation of probability density functions has been studied. Several known issues have been illustrated and elaborated. First, the presence of negative values of density approximations at many grid points has been identified as practically unavoidable, since it arises from the use of essential
(cheap and adaptive) techniques. The ratio of negative values can reach tens percent. Illustrations of the decomposition have been provided for grasping the visual intuition. Second, it has been documented how a correlation between a pair of state elements raises all ranks in between. In case of Gaussian densities and multiple correlated pairs, the ranks in the overlapping parts are almost multiplied. Last, the grid placement problem has been considered. Decorrelation does not guarantee reduction of tensor train ranks and any interpolation significantly limits the achievable precision.
  

\end{document}